\documentstyle[amssymb,aps,psfig]{revtex}

\newcommand{\bq}{\begin{equation}}
\newcommand{\eq}{\end{equation}}

\begin{document}

\twocolumn[\hsize\textwidth\columnwidth\hsize\csname
@twocolumnfalse\endcsname \draft \title{Orbital frustration at the
  origin of the magnetic behavior in LiNiO$_{2}$} \author{F.
  Reynaud$^{1}$, D. Mertz$^{1}$, F. Celestini$^{1}$, J.-M.
  Debierre$^{1}$,
  A. M.\ Ghorayeb$^{1}$, P.\ Simon$^{2}$, A. Stepanov$^{1}$, \\
  J. Voiron$^{3}$, and C.\ Delmas$^{4}$} \address{$^{1}$Laboratoire
  Mat\'{e}riaux et Micro\'{e}lectronique de Provence, Case 151,
  Universit\'{e}
  d'Aix-Marseille III and CNRS, \\13397 Marseille Cedex 20, France\\
  $^{2}$Centre de Recherche sur les Mat\'{e}riaux \`{a} Hautes
  Temp\'{e}ratures, CNRS,
  45071 Orl\'{e}ans Cedex 2, France\\
  $^{3}$Laboratoire Louis N\'{e}el, CNRS, Bo\^{\i}te Postale 166, 38042 Grenoble Cedex 9, France\\
  $^{4}$Institut de Chimie de la Mati\`{e}re Condens\'{e}e de
  Bordeaux, Ecole Nationale Sup\'{e}rieure de Chimie et Physique de
  Bordeaux and CNRS, 33608 Pessac Cedex, France} \date{february 27, 2001}
\maketitle

\begin{abstract}
  We report on the ESR, magnetization and magnetic susceptibility
  measurements performed over a large temperature range, from $1.5$ to
  $750$ K, on high-quality stoichiometric LiNiO$_{2}$. We find that
  this compound displays two distinct temperature regions where its
  magnetic behavior is anomalous.  With the help of a statistical
  model based on the Kugel'-Khomskii Hamiltonian, we show that below
  $T_{of}$ $\simeq 400$ K, an orbitally-frustrated state
  characteristic of the triangular lattice is established. This then
  gives a solution to the long-standing controversial problem of the
  magnetic behavior in LiNiO$_{2}$.
\end{abstract}

\pacs{PACS numbers: 75.40.Cx, 76.30.-v, 75.30.Et, 75.40.Mg }

\vskip 1pc

] \ \ \ \ \ \ \ 

\narrowtext The problem of orbital degeneracy (OD) in transition-metal
oxides has attracted considerable interest recently, due to its
drastic effect on the magnetic, elastic, transport and other
properties of these compounds \cite {KKmod}. Among the most studied
oxides exhibiting OD are manganites, vanadates, some titanates and
other systems. Here we discuss a new interesting aspect of OD: the
frustration of degenerate orbitals, as applied to lithium nickel
oxide.

Since LiNiO$_{2}$ was first studied as an ionic conductor \cite{Good},
the magnetic properties of this layered compound have been reported in
numerous publications\ but are still the subject of considerable
controversy. Quantum spin liquid \cite{Hira}, spin glass
\cite{Reim,Kanno1,Corti,Baj}, frustrated antiferromagnetism
\cite{Reim,Hiro,Roug1,Barra1}, quantum disordered state without a spin
gap \cite{Kanno2} and other scenarios \cite
{Kemp,Gang,Stoy,Rosen,Azzoni,Barra2,Mertz,Rey,Nun} have been claimed
to be responsible for the magnetism of this material. Actually,
interest in this compound stems essentially from two facts. First,
following the suggestion by Anderson \cite{Anderson} that a resonating
valence bond ground state may exist in a S=1/2 Heisenberg
triangular-lattice antiferromagnet, LiNiO$_{2}$ was proposed by
Hirakawa {\it et al.} \cite{Hira} as a potential physical realization
of such a system. The second important feature, overlooked in earlier
studies, is the unusual electronic state of Ni$^{3+}$ ions. The strong
crystal field acting on the nickel ions brings them in the low-spin
state $(t_{2g}^{6}$, $e_{g}^{1}$; $S=1/2)$, with the two $e_{g}$
levels being degenerate. Clearly, OD should play a crucial role in the
magnetic properties of this compound \cite{Kanno2,Feiner,Bossche}.
However, no direct confirmation of this hypothesis exists yet.

In this paper we propose a novel scenario for LiNiO$_{2}$ based on an
interplay between frustration on a triangular lattice and OD of
Ni$^{3+}$ ions. We show experimentally that an anomalous magnetic
behavior occurs in two different temperature regions, which we
interpret as the consequence of two distinct energy scales present in
the system. The first one, of the order of hundreds of Kelvin, is
related to the antiferro-orbital coupling between Ni$^{3+}$ ions and
leads to a frustrated distribution of Ni$^{3+}$ orbitals. The second
one, of the order of tens of Kelvin, is due to the spin-spin
interactions between electrons in these frustrated Ni$^{3+}$ orbitals,
and drives the very unusual magnetic behavior at low temperatures.
Our proposed scenario, which we confirm through a direct comparison
between experimental data and simulation results, gives a coherent and
natural explanation of this behavior.

High-quality powder samples of stoichiometric LiNiO$_{2}$ have been
prepared following the procedure described by Rougier {\it et
  al.}\cite{Roug2}. We have studied these samples by X-band ($\nu
=9.45\;$GHz) electron spin resonance (ESR), magnetization (up to
$15\;$T) and dc-susceptibility measurements over a wide temperature
range, from $1.5$ to $750\,$K.

The ESR spectra have a symmetric shape at all temperatures and show no
trace of any particular feature that would be representative of
ferrimagnetic cluster formation or of any anisotropy. This is a first
but clear indication that the samples studied in this work are
homogeneous and of good quality.  We show in Fig. 1 the $T$-dependence
of the ESR linewidth, $\Delta H$ (half width at half maximum), of a
typical sample of LiNiO$_{2}$ in the low-$T$ regime. A huge increase
of $\Delta H$ is noted when the temperature decreases below $50\,$K.\ 
This temperature, to which we shall refer below as
$T_{m}$, quite naturally gives the scale of the magnetic interaction in LiNiO%
$_{2}$. Below $T\simeq 10\,$K, $\Delta H$ tends to a saturation value
of about $0.5\,$T. The inset of Fig. 1 shows the inverse of the spin
susceptibility, $\chi ^{-1}(T)$, deduced from the ESR spectra. As
already
reported earlier \cite{Kanno1}, $\chi (T)$ follows a Curie-Weiss behavior, $%
\chi =C/(T-\theta )$, down to about $80$ K, with $\theta \simeq 35\pm
5$ K.  This low value of $\theta $ confirms that the sample studied is
so close to stoichiometry that we may consider it as pure LiNiO$_{2}$
\cite {Kanno1,Kanno2}. Bearing in mind that $\theta $ represents the
average sum of exchange interactions on an atomic site, the positive
sign of $\theta $ implies that the dominant interactions are
ferromagnetic (FM). However, the line broadening observed in Fig. 1 is
rather indicative of strong antiferromagnetic (AF) fluctuations. Line
narrowing would be expected if the
interactions were purely ferromagnetic. Futhermore, between $13\,$K and $%
50\, $K, $\Delta H$ can be fitted to the empirical formula $\Delta
H=AT^{-p}$ with $p=2.65$. Such an exponent value is known to be
characteristic of AF materials having a strong 2D character
\cite{Wijn}. Another important point is the tendency of the linewidth
to saturate below $10\,$K. This constitutes strong evidence that the
AF correlations do not propagate any longer below this temperature.

Fig. 2 shows the magnetic moment, $M$, as a function of magnetic
field, $H$.  The curves deviate from linearity below $T_{m}\simeq
50\,$K, the same temperature below which the ESR linewidth was seen to
markedly increase. At low temperatures, saturation is almost reached
in fields of $15\,$T and, furthermore, below about $10\,$K,
temperature has hardly any effect on magnetization. Although the
initial susceptibility increases with decreasing temperature in a
manner reminiscent of a FM behavior, the $M(H)$ curve at the lowest
temperature remains below what would be given by a Brillouin function.
This confirms the existence of rather strong AF interactions along
with ferromagnetism.

The second anomalous aspect of LiNiO$_{2}$ is found at high
temperatures and plays a crucial role in understanding the magnetic
properties of this compound. Fig. 3a shows the results of the
dc-susceptibility measurements, plotted in the form $(T+10)\chi (T)$
between $300\,$K and $750\,$K. The
curve presents a weak anomaly at around $400\,$K. At higher temperatures, $%
\theta $, which was found to be around $35\,$K from the low-$T$
measurements, decreases to finally settle at the negative value,
$\theta =-10\pm 5\,$K, for $T\geqslant \ 600\ $K. The inset of Fig. 3a
details the variation of $\theta (T)$, as deduced from $\chi (T)$.
Such a drastic change in $\theta $ is quite unusual and gives strong
evidence for a concomitant modification of the Ni$^{3+}$ orbital
distribution in the NiO$_{2}$ layers.  Additional support for this
argument comes from the ESR data. Fig.\ 3b presents the $T$-dependence
of $\Delta H$, between $300\,$K and $625\,$K.  The divergence of
$\Delta H$ above room temperature is quite in line with what is
expected when, due to changes in the Ni$^{3+}$ orbital distribution,
strong fluctuations occur. \ 

In order to explain this extremely rich and unusual magnetic behavior,
we use the Kugel'-Khomskii (KK) spin-orbital model\cite{KKmod}.
Derived from the two-band Hubbard Hamiltonian in the limit of a large
Coulomb repulsion, this model has proven its efficiency to describe
the interplay between magnetic and orbital ordering \cite{Pati}. Here
we rewrite the KK Hamiltonian in the following form :
\begin{equation}
H=\sum_{<i,j>}\left\{ -J^{o}\tau _{i}\tau _{j}-{\bf s}_{i}{\bf s}_{j}\left[
J_{so}\delta (\tau _{i}\tau _{j}-1)+J_{do}\delta (\tau _{i}\tau _{j}+1)%
\right] \right\}
\end{equation}
where the sum runs over nearest neighbor (NN) atoms and $\delta (\tau
_{i}\tau _{j}\pm 1)$ is equal to $1$ if $\tau _{i}\tau _{j}\pm 1=0$ and to $%
0 $ otherwise. As usual, ${\bf s}_{i}$ is the spin on atomic site $i$
while the pseudo-spin, $\tau _{i}$, is introduced to describe the
orbital occupancy. The values $\tau _{i}=+1$ and $\tau _{i}=-1$
respectively correspond to the two degenerate ${\em {e_{g}}}$ orbitals
of the Ni$^{3+}$ ion, $\left| x^{2}-y^{2}\right\rangle $ and $\left|
  3z^{2}-r^{2}\right\rangle $. In the first term, $J^{o}$ is the
orbital coupling constant. $J_{so}$ and $J_{do}$ are the magnetic
exchange constants between NN, respectively with the same and with a
different orbital occupancy. The spin-orbital model is here taken in
its simplest form. First, we consider ${\bf s}_{i}$ as classical Ising
spins. As a consequence, possible quantum effects at low temperatures
cannot be described. Second, we do not distinguish between different
relative orientations of the same orbital in the same site, as it
could be done using a Potts model. Finally,
we assume that the orbital coupling between two ions with the same orbital $%
\tau _{i}=\tau _{j}=1$ is the same as the one between two ions with
$\tau _{i}=\tau _{j}=-1$. These are rather crude approximations but we
shall see that this minimal model is apparently able to capture the
essential physics of the system, at least for temperatures greater
than $\simeq 5$\thinspace K.

Recently, van den Bossche, Zhang and Mila \cite{Bossche} have
suggested that the $SU(4)$ symmetry ($J_{so}=J_{do}=J^{o}$) could be
at the origin of the magnetic properties of LiNiO$_{2}$. Our
high-temperature experimental data presented above suggest that
$|J^{o}|>>|J_{so}|,\,|J_{do}|$. Therefore, we are here testing the
relevance of the alternative $SU(2)\ast SU(2)$ symmetry in which the
magnetic and orbital interactions have different orders of magnitude.

Concerning the sign of the orbital coupling, a ferro-orbital constant ($%
J^{o}>0$) would lead to a structural ferrodistortive phase transition.
This possibility is completely ruled out by the available X-ray
data\cite{Delmas}
so that we have to consider an antiferro-orbital coupling constant ($J^{o}<0$%
). Since the Ni$^{3+}$ ions form a triangular arrangement, we then
expect
{\it orbital frustration} to occur below a temperature $T_{of}$ ($T_{of}$ $%
=(3/2)\left| J^{o}\right| /k_{B}$, using the mean-field approximation)
in a manner similar to the well-known AF-spin-frustration phenomenon.
From the
experimental results shown in Fig.\ 3, $T_{of}$\thinspace\ $\simeq \,400\,$%
K, which gives $\left| J^{o}\right| /k_{B}\simeq 270\,$K. The
orbitally-frustrated state may be described as a Wannier state (WS)
\cite {Wannier}. A particular property of a WS is that three NN cannot
have the same orbital occupancy. This is illustrated in Fig. 4 (a and
b) where we represent typical orbital distributions above and below
$T_{of}$. It is evident that at high temperatures the probability to
find a link between NN
with a different orbital occupancy is $1/2$ and that this probability is $%
2/3 $ in the WS. The mean-field values of $\theta $ are then given by:
\[
\theta _{ht}=(3J_{do}+3J_{so})/k_{B}\text{ and }\theta
_{ws}=(4J_{do}+2J_{so})/k_{B}
\]
where $\theta _{ht}$ and $\theta _{ws}$ denote the values in the
high-temperature state and in the WS, respectively. The assumption of
AF-orbital coupling, together with the experimentally inferred $\theta
_{ht}=-10\pm 5$ K and $\theta _{ws}=35\pm 5$ K, allows us to determine
the two magnetic coupling constants, $J_{do}$ and $J_{so}$. We find a
{\it FM exchange between NN with different orbitals}
($J_{do}/k_{B}=18\pm 6\,$K) and
an {\it AF one between NN with similar orbitals} ($J_{so}/k_{B}=-20\pm 6\,$%
K). It is interesting to note that this simple treatment of
Hamiltonian (1) is sufficient for the qualitative understanding of the
magnetic behavior of LiNiO$_{2}$\ in a large temperature interval
($T>T_{m}$). It can also be shown that these values of $J_{do}$ and
$J_{so}$\ constitute the only combination of magnetic coupling
constants that ensures the absence of a trivial long-range order, as
indicated by the saturation of $\Delta H(T)$ at low temperatures
(Fig.\ 1).

To test the ability of the model to describe the low-temperature
magnetic properties, we have performed Monte-Carlo (MC) simulations.
The simulated system contains $N_{i}$ = $36\times $ $36$ coplanar ions
and the classical Metropolis algorithm \cite{Metro} is used. At each
step both a spin and a pseudo-spin are chosen randomly and attempts to
flip them are performed independently. A MC step consists of $N_{i}$
such attempts. After a first run of $2\times $ $10^{4}$ MC steps to
ensure equilibrium, the magnetization and susceptibility are computed
from uncorrelated configurations recorded during a second run of
$5\times $ $10^{4}$ MC steps. The best fit to the
experimental results is obtained for $J_{do}/k_{B}=21.0\,$K and $%
J_{so}/k_{B}=-22.5\,$K. Considering the simplicity of the model, the
slight difference between these values and those deduced
experimentally is not surprising. The good overall agreement found
with the experimental data is rather satisfactory (Figs. 1 and 2) and
allows us to interpret the magnetic
behavior at intermediate and low temperatures. In the range $T_{m}<T<T_{of}$%
, the magnetic state is a paramagnetic one with $\theta >0$. There are
two thirds of NN pairs with different orbitals and hence a majority of
FM exchanges ($J_{do}>0$) so that the system behaves as if approaching
a FM transition. Since $|J_{so}|>|J_{do}|$, at lower temperatures the
links between the NN having identical orbitals are preferably
satisfied (see Fig.  4c), inducing AF-like fluctuations because
$J_{so}<0$. This situation finally leads to a{\it \ magnetically
  frustrated }state in LiNiO$_{2}$ at low temperatures.

To conclude, an important outcome of our experimental results is that
there
are two distinct energy scales characteristic of the magnetism of LiNiO$_{2}$%
, which respectively correspond to the antiferro-orbital coupling between Ni$%
^{3+}$ ions in the NiO$_{2}$ layers and to their spin-spin
interactions. The adjunction of two facts, orbital degeneracy of the
Ni$^{3+}$ ions, and their triangular arrangement, leads to the
build-up of a Wannier orbitally-frustrated state below $T_{of}\simeq
400\,$K. This uncommon orbital state is at the origin of the observed
low-$T$ complex magnetic behavior of LiNiO$_{2}$.

We would like to thank D.\ I.\ Khomskii, Y.\ Ksari, F.\ Mila and F.-C.
Zhang for fruitful discussions and V. A. Pashchenko for assistance in
the magnetic
measurements. 

\bigskip

\begin{figure}
  \psfig{file=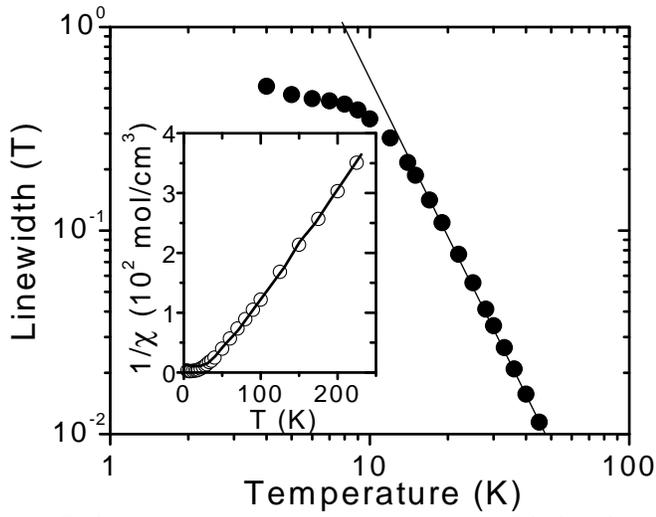,width=8.6cm,clip=t}
\caption{Temperature dependence of the ESR linewidth, $\Delta H$.\ The
  straight line is a fit to the law $\Delta H=AT^{-p}$ with $p=2.65$.
  The inset shows the temperature dependence of the inverse of the
  spin susceptibility, $1/\protect\chi $, as deduced from the ESR
  spectra. Circles and continuous line respectively represent
  experimental data and numerical simulations.}
\end{figure}
\bigskip \bigskip \bigskip

\begin{figure}
  \psfig{file=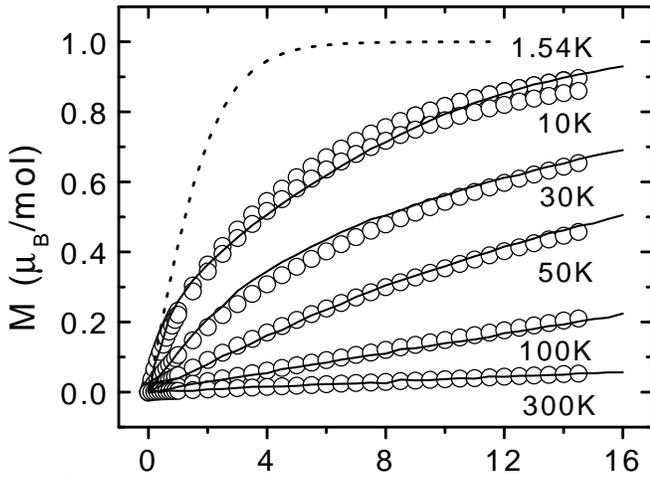,width=8.6cm,clip=t}
\caption{Magnetic moment, $M$, as a function of magnetic field, $H$, up to $%
15\,$T. Circles: experimental data between $300$\thinspace K and $1.54$%
\thinspace K; continuous curves: simulation results; dotted curve:
$M(H)$ calculated at $1.5\,$K using the Brillouin function.}
\end{figure}
\bigskip

\begin{figure}
  \vbox{ \LARGE\bf{a)} \vskip 1pc
    \psfig{file=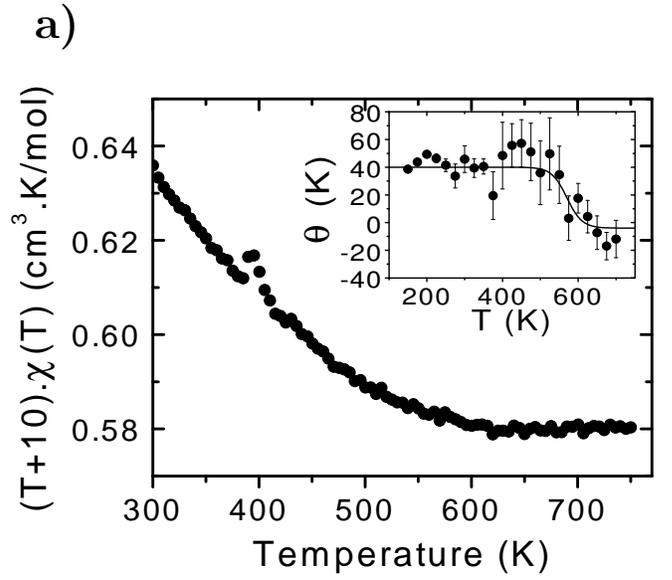,width=8.6cm,clip=t} \LARGE\bf{b)}
    \psfig{file=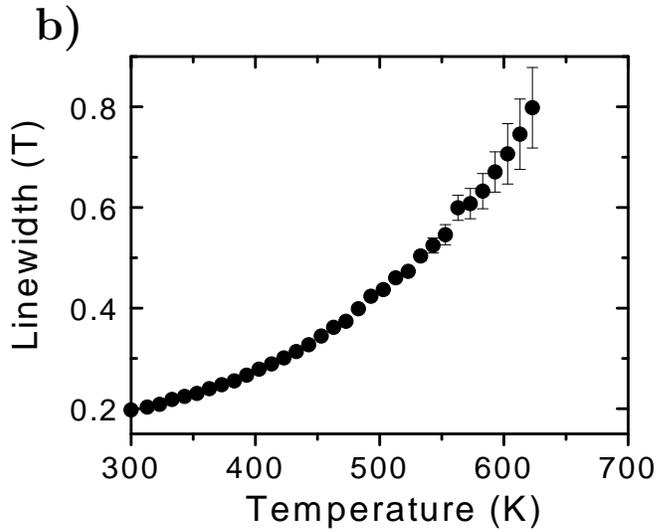,width=8.6cm,clip=t} }
\caption{a) Results of dc-susceptibility measurements, plotted in the form $%
  (T+10)\protect\chi (T)$. Note the plateau for $T>600\,$K. The inset
  shows the variation of $\protect\theta $ with temperature; the line
  is a guide to
the eye. b) Temperature dependence of $\Delta H$ between $300\,$K and $625\,$%
K.}
\label{Fig. 3}
\end{figure}

\bigskip

\begin{figure}
  \psfig{file=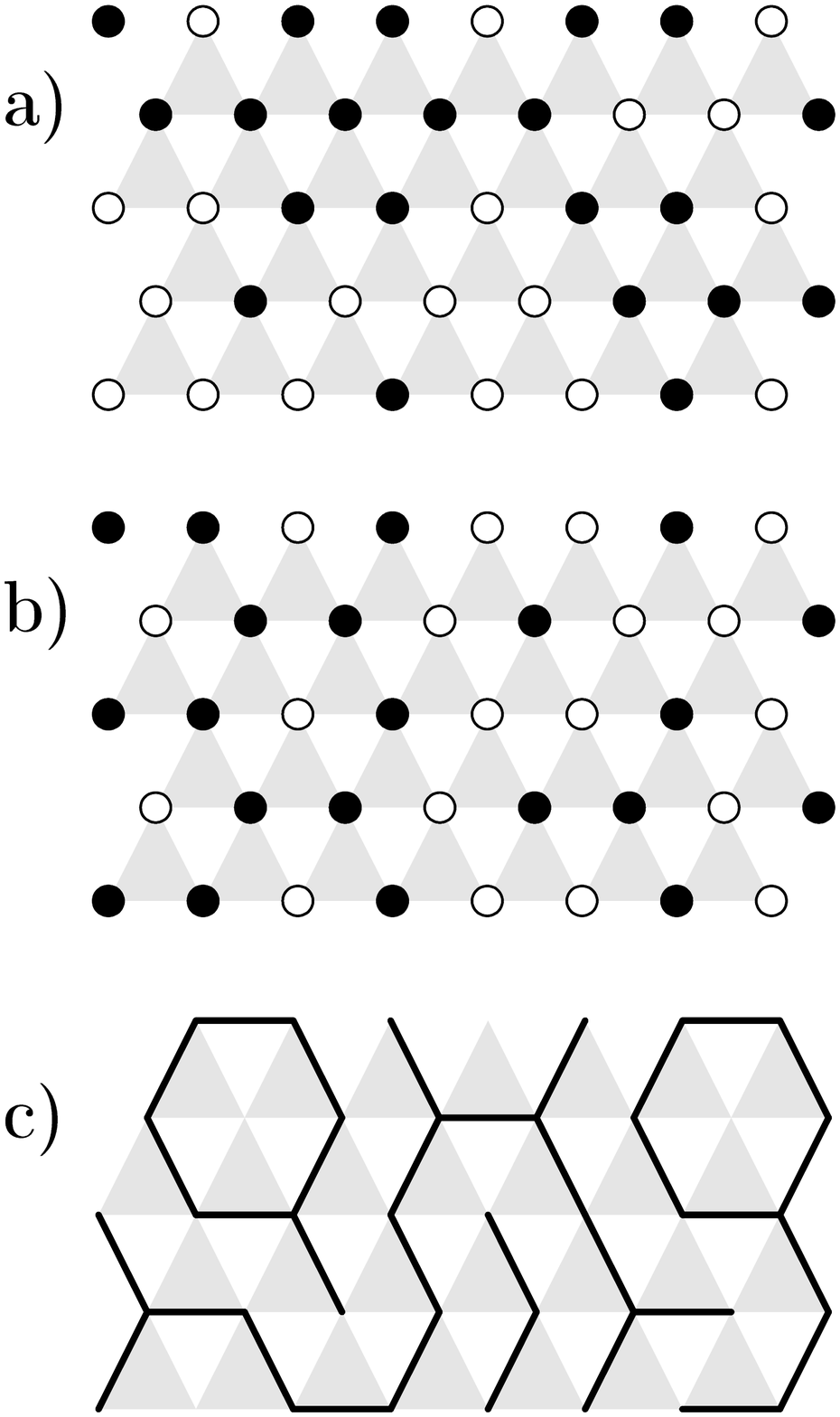,width=8.6cm,clip=t}
\caption{Simulated distribution of orbitals at high (a) and low (b)
  temperatures. Filled and open circles distinguish between the two
  types of orbitals. c) Satisfied FM links in the low-T Wannier state
  represented in (b).}
\end{figure}


\begin{references}
\bibitem{KKmod} K.\ I.\ Kugel' and D.\ I.\ Khomskii, Sov. Phys. -
  JETP\ {\bf 37}, 725 (1973); Usp.\ Fiz.\ Nauk {\bf 136}, 621 (1982)
  [Sov. Phys. - Usp. {\bf 25}, 231 (1982)].
  
\bibitem{Good} J.B.\ Goodenough, D.\ G.\ Wickham and W.\ J.\ Croft,
  J.\ Phys.\ Chem. Sol. {\bf 5}, 107 (1958).
  
\bibitem{Hira} K. Hirakawa, H.\ Kadowaki and K.\ Ubukoshi, J. Phys.
  Soc.
Jpn {\bf 54}, 3526 (1985); K. Hirakawa {\it et al.}, J. Phys. Soc. Jpn {\bf %
  59}, 3081 (1990) and references therein.

\bibitem{Reim} J. N. Reimers {\it et al.}, J. Sol. St. Chem. {\bf
    102}, 542 (1993).
  
\bibitem{Kanno1} K. Yamaura {\it et al.}, J. Sol. St. Chem. {\bf 127},
  109 (1996).
  
\bibitem{Corti} M. Corti {\it et al.}, J. Appl. Phys. {\bf 79}, 6621
  (1996).
  
\bibitem{Baj} A. Bajpai and A. Banerjee, Phys. Rev. B {\bf 55}, 12439
  (1997).
  
\bibitem{Hiro} H. Yoshizawa {\it et al.}, J. Phys. Soc. Jpn {\bf 59},
  2631
(1990); K. Hirota, Y.\ Nakazawa and M.\ Ishikawa, J. Mag. Mag. Mat. {\bf %
  90-91}, 279 (1990); J. Phys. : Cond. Mat. {\bf 3}, 4721 (1991).

\bibitem{Roug1} A. Rougier, C.\ Delmas and G.\ Chouteau, J. Phys.
  Chem.  Sol. {\bf 57}, 1101 (1996).
  
\bibitem{Barra1} A.-L. Barra {\it et al.}, J. Mag. Mag. Mat. {\bf
    177-181}, 783 (1998).
  
\bibitem{Kanno2} Y.\ Kitaoka {\it et al.}, J. Phys. Soc. Jpn {\bf 67},
  3703 (1998).
  
\bibitem{Kemp} J. P. Kemp, P.\ A.\ Cox and J.\ W.\ Hodby, J. Phys. :
  Cond.  Mat. {\bf 2}, 6699 (1990).
  
\bibitem{Gang} P.\ Ganguly {\it et al.}, Phys.\ Rev.\ B {\bf 46},
  11595 (1992).
  
\bibitem{Stoy} R. Stoyanova, E.\ Zhecheva and C.\ Friebel, J. Phys.
  Chem.  Sol. {\bf 54}, 9 (1993).
  
\bibitem{Rosen} M. Rosenberg {\it et al.}, J. Appl. Phys. {\bf 75},
  6813 (1994).
  
\bibitem{Azzoni} C.\ B.\ Azzoni {\it et al.}, Phys.\ Rev.\ B {\bf 53},
  703 (1996).
  
\bibitem{Barra2} A.-L. Barra {\it et al.}, Eur.\ Phys. J.\ B {\bf 7},
  551 (1999).
  
\bibitem{Mertz} D. Mertz {\it et al.}, Phys.\ Rev.\ B {\bf 61}, 1240
  (2000).
  
\bibitem{Rey} F.\ Reynaud {\it et al.}, Eur.\ Phys. J.\ B {\bf 14}, 83
  (2000).
  
\bibitem{Nun} M.\ D.\ N\'{u}\~{n}ez-Regueiro {\it et al.}, Eur.\ Phys.
  J.\ B {\bf 16}, 37 (2000).
  
\bibitem{Anderson} P.\ W.\ Anderson, Mater. Res.\ Bull. {\bf 8}, 153
  (1973).
  
\bibitem{Feiner} L.\ F.\ Feiner, A.\ M.\ Ole\'{s} and J.\ Zaanen,
  Phys.  Rev. Lett. {\bf 14}, 2799 (1997).
  
\bibitem{Bossche} M. van den Bossche, F.-C. Zhang and F.\ Mila, Eur.
  Phys.\ J.\ B {\bf 17}, 367 (2000).
  
\bibitem{Roug2} A. Rougier, P.\ Gravereau and C.\ Delmas, J.
  Electrochem.  Soc. {\bf 14}, 1168 (1996).
  
\bibitem{Wijn} H.\ W.\ de Wijn {\it et al.}, Sol.\ St.\ Commun. {\bf
    11}, 803 (1972).
  
\bibitem{Pati} See, for example, S.\ K.\ Pati, R.\ R.\ P.\ Singh and
  D.\ I.\ Khomskii, Phys.\ Rev.\ Lett. {\bf 81}, 5406 (1998); G.\ 
  Santoro {\it et al.} Phys.\ Rev.\ Lett. {\bf 83}, 3065 (1999) and
  references therein.
  
\bibitem{Delmas} A.\ Rougier, C.\ Delmas and A.\ V.\ Chadwick, Sol.
  St.\ Commun. {\bf 94}, 123 (1995).
  
\bibitem{Wannier} G.\ H.\ Wannier, Phys.\ Rev.\ {\bf 79}, 357 (1950);
  Phys.\ Rev.\ B {\bf 7}, 5017 (1973).
  
\bibitem{Metro} N.\ Metropolis {\it et al.}, J.\ Chem.\ Phys. {\bf
    21}, 1087 (1953).
\end{references}
\end{document}